\documentclass{aastex}
\usepackage{spr-astr-addons}

\def\aap{A\&A\,  }
\def\apj{ApJ\,  }
\def\apjl{ApJ\,  }
\def\apjs{ApJS  }
\def\iaucirc{IAU circ.  } 
 
\def\mnras{MNRAS\,  }

\RequirePackage{color}

\def\snr{SN\,1993J~}
\def\sn2002bo{SN\,2002bo~}
\def\s2003dh{SN\,2003dh~}

\begin{document}

\title
{
Time-Dependent Models for  a decade of SN 1993J
}
\shorttitle{Models for SN 1993J}
\shortauthors{Zaninetti}

\author{L. Zaninetti \altaffilmark{1}} 
\affil{
Dipartimento di Fisica Generale,   \\
Universit\`a degli Studi di Torino \\
Via Pietro Giuria 1,               \\
           I-10125 Torino, Italy }

\begin{abstract}
A  classical and a relativistic  law of motion
for a  supernova remnant (SNR) are deduced
assuming an inverse power law behavior
for the density of the interstellar
medium and applying the thin layer approximation.
A third equation of motion is found
in the framework of relativistic hydrodynamics
with pressure, applying momentum conservation.
These new formulas are calibrated against a decade
of observations of \snr.
The existing knowledge of the diffusive processes
of ultrarelativistic electrons is
reviewed in order to explain the behavior
of the `U' shaped profile
of intensity versus distance from the
center of SN\,1993J.
\end{abstract}


\section{Introduction}

The study of
the supernova remnant (SNR)  started  with
\cite{Oort1946}
where an on ongoing  collisional excitation
as a result  of a post-explosion
expansion of the SNR  against the
ambient medium was suggested.
The next six decades
where dedicated
to the deduction of an analytical or numerical
law of expansion.
The target  is  a relationship  for the  instantaneous
radius of expansion, $R$,
of the type
$\propto~ t^m $ where $t$  is time
and $m$ is a parameter that depends on the chosen model.
On adopting  this point of view,  the Sedov expansion
predicts  $R \propto  t^{0.4}$, see \cite{Sedov1959},
and the thin layer approximation  in the
presence of a constant density medium
predicts $R \propto  t^{0.25}$,
see \cite{Dyson1997}.
A  simple approach  to the SNR evolution
in the first $10^4$ yr
assumes an initial  free expansion in  which
$R \propto t$  until the surrounding mass
is of the order of  1 $M_{\sun}$ and a second phase
characterized by the energy conservation in  which according
to the Sedov solution $R \propto t^{2/5}$,
see \cite{Dalgarno1987}.
A third phase characterized by an adiabatic expansion
with $R \propto t^{2/7}$
starts after $10^4$ yr, see \cite{Dalgarno1987}.
A more sophisticated approach given
by \cite{Chevalier1982a,Chevalier1982b}
analyzes  self-similar solutions
with varying inverse power law exponents
for the density profile of the advancing matter,
$R^{-n}$, and ambient  medium,
$R^{-s}$. The  previous assumptions give
a law  of motion
$R \propto  t^{\frac{n-3}{n-s} }$  when
 $n \, > 5$.
Another example  is an analytical solution suggested
by  \cite{Truelove1999} where the radius--time relationship
is regulated by the decrease in density:
as an example, a density  proportional to $R^{-9}$
gives  $R\propto t ^{2/3}$.
With regard to observations,
the radius--time relationship was clarified
when  a decade of
very-long-baseline interferometry (VLBI)
observations of \snr at wavelengths of
3.6, 6, and 18 cm  became available,
see  \cite{Marcaide2009}.
As a first example, these observations
collected over a 10 year period
can be approximated by  a power law dependence
of the type  $R\,  \propto  t^{0.82}$.
This observational fact  rules  out the
Sedov model and the momentum conservation model.
The observed  radius--time relationship
leaves a series of questions
unanswered or merely partially answered:
\begin{itemize}
\item
Is it possible to deduce  a classical  equation
of motion for the SNR with an adjustable
parameter  that can be found from  a numerical  analysis
of the radius--time relationship?
\item
Is it possible to deduce a relativistic-mechanics equation
of motion for the SNR, since
the initial velocity of the SNR can be on
the order of 1/3  of the velocity of light?
\item
Is it possible to deduce a relativistic-hydrodynamics equation
of motion for the SNR
applying momentum conservation?
\item
Can we build a diffusive model
which  explains  the behavior of the
emission intensity of the SNR?
\item
Can a simple model  for the time evolution
of the total mapped flux
densities of \snr be built?
\end{itemize}

In order to answer  these questions, Section \ref{sec_data}
reports the data on \snr.
Section \ref{sec_classical} reports a new classical
law of motion assuming an inverse
 power  law  dependence
for the density of the medium.
Section \ref{sec_fits}  reports the evolution of \snr
on the basis of four models.
Section \ref{sec_relativistic}  contains two new
relativistic laws of motion as well as a fit to the data.
Section \ref{sec_transfer} reviews the existing situation
with the radiative transport equation as well as three  different
processes of diffusion for ultrarelativistic electrons.
Section \ref{sec_transfer} also contains a new
scaling law for the temporal evolution of the flux of
\snr at 6 cm.

\section{A spherical SNR}

\label{sec_data}
\begin{figure*}
\begin{center}
\includegraphics[width=7cm ]{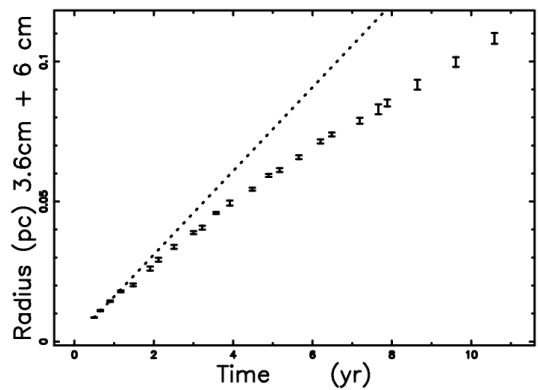}
\end {center}
\caption
{
Radius in  pc versus year of \snr with
vertical error bars.
The em bands  are  $\lambda=3.6 \, cm $
and  $\lambda=6 \, cm $.
The data are extracted from Table 1 in Marcaide et al. 2009.
The dotted line represents an expansion at a
constant velocity.
}
\label{1993pc}
    \end{figure*}

The supernova \snr started to be visible
in M81 in 1993, see \cite{Ripero1993},
and presented a
circular symmetry
for 4000
days, see \cite{Marcaide2009}.
Its distance is
3.63~Mpc (the same as M81), see \cite{Freedman1994}.
The expansion of \snr  has been monitored in various bands
over a decade  and Fig. \ref{1993pc} reports
its temporal evolution.

\begin{figure*}
\begin{center}
\includegraphics[width=7cm ]{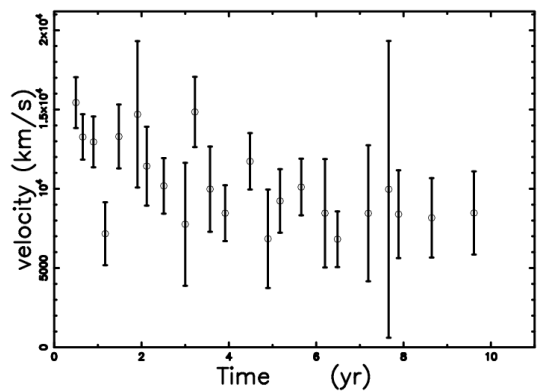}
\end {center}
\caption
{
Instantaneous velocity
of \snr with
uncertainty.
}
\label{1993realvel}
    \end{figure*}

The instantaneous velocity of expansion
can be deduced from the following formula
\begin{equation}
v_j = \frac
{r_{j+1} - r_j}
{t_{j+1} - t_j}
\quad ,
\label{vdiscrete}
\end{equation}
where $r_j$ and $t_j$  denote the radius  and the time
at the position $j$.
The uncertainty in the instantaneous   velocity
is found by implementing the error
propagation equation (often called the law of errors of Gauss) when
the covariant terms are neglected, see ~\cite{Bevington2003}).
Fig. \ref{1993realvel} reports the instantaneous velocity
aw well the relative uncertainty.
In particular, the observed instantaneous velocity decreases from
$v=15437~\frac{km}{s}$ at $t=0.052$\ yr  to
$v=8474~ \frac{km}{s}$ at $t=10.53$\ yr.
We briefly recall that \cite{Fransson2005}
quote
an inner velocity
from  the shapes of the lines
of  $\approx 7000 \frac{km}{s}$
and
an outer  velocity of $\approx 10000 \frac{km}{s}$.

\section{Classical case}
This section
reviews
the free expansion ,
two simple laws of
motion for the SNR and a new law of motion
in the light of the classical physics.

\subsection{The constant expansion velocity}

The SNR  expands at a constant velocity until
the surrounding mass is
of the order of the solar mass.
This time , $t_M$ ,
is
\begin {equation}
t_M= 186.45\,{\frac {\sqrt [3]{{\it M_{\sun} }}}{\sqrt [3]{{\it n_0}}{\it
v_{10000}}}} \quad yr
\quad ,
\end{equation}
where $ M_{\sun}$ is the number of solar masses
in the volume occupied by the SNR,
$n_0$,  the
number density  expressed  in particles~$\mathrm{cm}^{-3}$,
and $v_{10000}$ the initial velocity expressed
in units of $10000\,km/s$ , see \cite{Dalgarno1987}.

\subsection{Two solutions}

\begin{figure*}
\begin{center}
\includegraphics[width=7cm ]{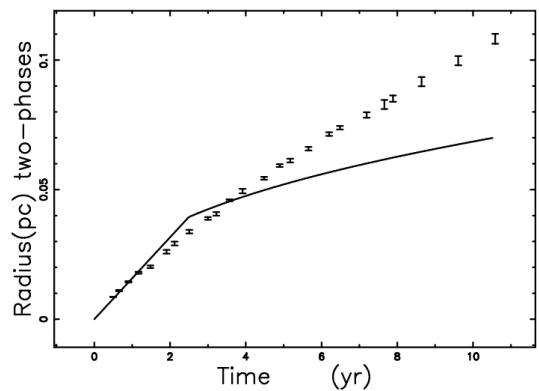}
\end {center}
\caption
{
Theoretical radius as given by the two-phases
solution
with data as in  Table~\ref{datafit} (full line),
and astronomical data of \snr with
vertical error bars.
}
\label{1993duefasi}
    \end{figure*}

A first law of motion for the $SNR$
is the  Sedov  solution
\begin{equation}
R(t)=
\left ({\frac {25}{4}}\,{\frac {{\it E}\,{t}^{2}}{\pi \,\rho}} \right )^{1/5}
\quad ,
\label{sedov}
\end{equation}
where $E$ is the energy injected into the process
and $t$ is  time,
see~\cite{Sedov1959,Dalgarno1987}.
Our astrophysical  units are: time, ($t_1$), which
is expressed  in years;
$E_{51}$, the  energy in  $10^{51}$ \mbox{erg};
$n_0$,  the
number density  expressed  in particles~$\mathrm{cm}^{-3}$~
(density~$\rho=n_0$m, where m = 1.4$m_{\mathrm {H}}$).
In these units, equation~(\ref{sedov}) becomes
\begin{equation}
R(t) \approx  0.313\,\sqrt [5]{{\frac {{\it E_{51}}\,{{\it t_1}}^{2}}{{\it n_0}}}
}~{pc}
\quad .
\end{equation}
The Sedov solution scales as $t^{0.4}$.
We are now ready to couple
the Sedov phase with the free expansion phase
\begin{equation}
 R(t)  = \left\{ \begin{array}{ll}
0.0157\,{\it t}\, pc &
\mbox {if $t \leq 2.5 yr$ } \\
0.0273\,\sqrt [5]{{{\it t}}^{2}} \, pc   &
\mbox {if $t >    2.5yr $ ~.}
            \end{array}
            \right.
\label{twophases}
\end{equation}
This two phases solution is obtained with
the following parameters
$M_{\sun}$ =1 ,
$n_0 =1.127 \,10^5$ ,
$E_{51}=0.567$
and Fig. \ref{1993duefasi} reports it's  temporal behavior
as well the data.
The quality  of the simulation  , $\epsilon_q$,
can be obtained
by  a comparison of  observed and simulated quantities:
\begin{equation}
\epsilon_q  =(1- \frac{\vert( R_{\mathrm {pc,obs}}- R_{pc,\mathrm {num}}) \vert}
{R_{pc,\mathrm {obs}}}) \cdot 100
\,,
\label{efficiency}
\end{equation}
where $R_{pc,\mathrm {obs}}$ is the
observed radius, in parsec
and $R_{pc,\mathrm {num}}$ is the radius  from our simulation
in parsec.
In the case of  the two-phases simulation
we have $\epsilon_q=65 \%$  which is a low value
in comparison with  the other models here considered
see  Table~\ref{datafit}.

A second solution is connected with
momentum conservation in the presence
of a constant density medium,
see \cite{Dyson1997,Padmanabhan_II_2001,Zaninetti2009a}.
The astrophysical radius in pc as a function
of time is
\begin{equation}
R(t) =
\sqrt [4]{{{\it R_{0}}}^{3} \left(
 4.08\,10^{-6}\,{\it v_{1}}\, \left( t_1-t_{{0}} \right) +{\it R_{0}
} \right) } \, pc
\quad ,
\end{equation}
where
$t_1$ and $t_0$ are  times in years,
$R_0$ is the  radius in pc  when  $t_1=t_0$ and
$v_{1}$ is the velocity in
$\frac{1\,km}{s}$ units  when
$t_1=t_0$.
The momentum solution in the presence of a constant density medium
scales as $t^{0.25}$.

\subsection{Momentum conservation with variable density}

\label{sec_classical}
We assume that around the SNR the density of the
interstellar medium (ISM)
has
the following two piecewise dependencies
\begin{equation}
 \rho (R)  = \left\{ \begin{array}{ll}
            \rho_0                      & \mbox {if $R \leq R_0 $ } \\
            \rho_0 (\frac{R_0}{R})^d    & \mbox {if $R >    R_0 $ ~.}
            \end{array}
            \right.
\label{piecewise}
\end{equation}
In this framework, the density decreases as
an inverse power law with an exponent $d$ that can be fixed from the
observed temporal evolution of the radius,
with $d=0$ meaning constant radius.
The mass swept, $M_0$,
in the interval $0 \leq r  \leq R_0$
is
\begin{equation}
M_0 =
\frac{4}{3}\,\rho_{{0}}\pi \,{R_{{0}}}^{3}
\quad .
\end{equation}
The mass swept, $ M $,
in the interval $0 \leq r \leq R$
with $r \ge R_0$
is
\begin{eqnarray}
M =
-4\,{r}^{3}\rho_{{0}}\pi \, \left( {\frac {R_{{0}}}{r}} \right) ^{d}
 \left(d -3 \right) ^{-1}   \nonumber \\
+4\,{\frac {\rho_{{0}}\pi \,{R_{{0}}}^{3}}{d-
3}}
+ \frac{4}{3}\,\rho_{{0}}\pi \,{R_{{0}}}^{3}
\quad .
\end{eqnarray}
Momentum conservation requires  that
\begin{equation}
M v = M_0 v_0
\quad ,
\end {equation}
where  $v$   is  the velocity at $t$
and    $v_0$ is  the velocity at $t=t_0$.
This  formula is not invariant
under Lorentz transformations and the initial
velocity, $v_0$, can be greater than the velocity of light.
The previous expression as a function of the radius
is
\begin{equation}
\beta =
\frac
{
{{\it r_0}}^{3}{\it \beta_0}\, \left( 3-d \right)
}
{
3\,{{\it r_0}}^{d}{R}^{3-d}-{{\it r_0}}^{3}d
}
\quad ,
\label{velclassic}
\end{equation}
where
$\beta_0$=$v_0/c_l$,
$\beta$=$v/c_l$
and $c_l$   is the velocity of light.
Here, we have introduced a relativistic notation
for later use.
In this differential equation of first order
in $R$, the
variables can be separated and an integration
term-by-term gives
the following
nonlinear equation ${\mathcal{F}}_{NL}$
\begin{eqnarray}
{\mathcal{F}}_{NL} =
 \left( 4{R_{{0}}}^{3}d-{R_{{0}}}^{3}{d}^{2} \right) R -3{R_{{0}}}^
{d}{R}^{4-d}+{R_{{0}}}^{4}{d}^{2}
\nonumber \\
+12{R_{{0}}}^{3}v_{{0}}t+3{R_{{0}
}}^{4}-4{R_{{0}}}^{4}d
\nonumber\\
+7{R_{{0}}}^{3}v_{{0}}d{\it t_0}+{R_{{0}}}^{3
}v_{{0}}{d}^{2}t
\nonumber \\
-7{R_{{0}}}^{3}v_{{0}}dt
-12{R_{{0}}}^{3}v_{{0}}{
\it t_0}-{R_{{0}}}^{3}v_{{0}}{d}^{2}{\it t_0}
=0
\quad  .
\label{nonlinear}
\end {eqnarray}
An approximate solution of
${\mathcal{F}}_{NL}(r) $  can be obtained
assuming that
$3 R_0^d R^{4-d}$
$\gg$
$-(4 R_0^3 d-R_0^3 d^2)R$
\begin{eqnarray}
 R(t) = \nonumber \\
 ( {R_{{0}}}^{4-d}-\frac{1}{3}d{R_{{0}}}^{4-d}
( 4-d ) \nonumber \\
 + \frac{1}{3}
 ( 4-d ) v_{{0}}{R_{{0}}}^{3-d} ( 3-d )
 ( t-t_{{0}} )  ) ^{\frac{1}{4-d}}
\quad .
\label{asymptotic}
\end{eqnarray}
Up to now, the physical units have not been specified,
pc for length  and  yr for time
are perhaps an acceptable choice.
With these units, the initial velocity $v_{{0}}$
is  expressed in $\frac{pc}{yr}$ and should be converted
into $\frac{km}{s}$; this means
that $v_{{0}} =1.02\,10^{-6} v_{{1}}$
where  $v_{{1}}$ is the initial velocity expressed in
$\frac{km}{s}$.

The astrophysical version of the above equation
in pc is
\begin{eqnarray}
 R(t) = \nonumber \\
 ( {R_{0}}^{4-d}-\frac{1}{3}d{R_{{0}}}^{4-d}
( 4-d ) 
 +3.402 \,10^{-7} \times \nonumber \\
 \times( 4-d ) v_{{1}}{R_{{0}}}^{3-d} ( 3-d )
 ( t_1-t_{{0}} )  ) ^{\frac{1}{4-d}} \, pc
\quad ,
\label{radiusvarpc}
\end{eqnarray}
where
$t_1$ and $t_0$ are  times in  years,
$R_0$ is the radius in pc  at $t_1=t_0$ and
$v_{1}$ is the velocity at
$t_1=t_0$
in $\frac{km}{s}$.

\section{Classical fits}
\label{sec_fits}
The quality of the fits is measured by the
merit function
$\chi^2$
\begin{equation}
\chi^2  =
\sum_j \frac {(R_{th} -R_{obs})^2}
             {\sigma_{obs}^2}
\quad ,
\label{chisquare}
\end{equation}
where  $R_{th}$, $R_{obs}$ and $\sigma_{obs}$
are the theoretical radius, the observed radius and
the observed uncertainty respectively.

A {\em first} numerical analysis of the observed
radius--time relationship of \snr
can be done by assuming  a  power law
dependence  of the type
\begin{equation}
R(t) = r_p t^{\alpha_p}
\label{rpower}
\quad .
\end{equation}
The two parameters $r_p$ and  $\alpha_p$ can be found
from the following
logarithmic transformation
\begin{equation}
\ln(R(t)) = \ln(r_p) + \alpha_p \ln (t)
\quad,
\end{equation}
which can be written as
\begin{equation}
y  = a_{LS} +b_{LS}x
\quad  .
\end{equation}
The application
of the least square method
through the FORTRAN subroutine LFIT from
\cite{press} allows to find $a_{LS}$ ,$b_{LS}$ and the
errors $\sigma_a$ and $\sigma_b$,
see numerical values in
Table~\ref{datafit}.
The error on $r_p$
is  found by implementing the error
propagation equation
(often called law of errors of Gauss) when
the covariant terms are neglected
(see  equation (3.14)
in~\cite{Bevington2003}),
\begin{equation}
\sigma_{r_p}  =\exp (a) \sigma_a
\quad  ,
\end{equation}
and $\sigma_{\alpha_p}$ = $\sigma_b$.
In this case, the velocity is
\begin{equation}
V(t) = r_p \, \alpha_p t^{(\alpha_p-1)}
\quad .
\label {vpower}
\end{equation}

\begin{table}
\caption { Numerical values of the parameters of the fits and
$\chi^2$. $N$  represents the number of  free parameters and
$\epsilon_q$ the quality of the simulation.
 }
 \label{datafit}
 \[
 \begin{array}{cccc}
 \hline
 \hline
 \noalign{\smallskip}
  N
& values  & \chi^2      & \epsilon_q   \\
 \noalign{\smallskip}
 \hline
 \noalign{\smallskip}
  &power~law   &   &\\ \noalign{\smallskip}
  2   & \alpha_p = 0.82 \pm 0.0048  & 6364 & 98.54 \%
\\ \noalign{\smallskip}
 ~    &r_p = (0.015 \pm 0.00011)  & &  \\
 \noalign{\smallskip}
 \hline
        & piecewise   &   &\\ \noalign{\smallskip}
 4   & \alpha_1 = 0.83 \pm 0.01  & 32
& 97.76 \%
 \\
   & \alpha_2 = 0.78 \pm 0.0077; & ~
& ~
 \\

 \noalign{\smallskip}
 ~   & r_{br} = 0.05~{pc};
 t_{br}=4.10~{yr}
& ~ &~
 \\
 \noalign{\smallskip}
 \hline
  &approximate~radius  &   &\\ \noalign{\smallskip}
  4   & d=2.54; r_{0} = 0.019~{pc}; &  7186 &
96.95 \%
\\
\noalign{\smallskip} ~   & t_{0}=0.249~{yr}; v_0 =100 000
\frac{km}{s} & ~ & ~
\\
\noalign{\smallskip} \hline
~ &  nonlinear~radius  &   &\\ \noalign{\smallskip}
  4   &
d=2.93; r_{0} = 0.019~{pc}; &   276 & 93.3 \%
 \\\noalign{\smallskip}
   & t_{0}=0.249~{yr}; v_0 =100 000
\frac{km}{s} & ~ &~
\\
\noalign{\smallskip}
 \hline
      &  relativistic~radius &   &\\ \noalign{\smallskip}
4   & d=2.54; r_{0} = 0.0045~{pc}; &  5557 & 93.05 \%
\\
\noalign{\smallskip}  ~   & t_{0}=0.249~{yr}; \beta_0= 0.333 & ~
&~
\\
\noalign{\smallskip}
 \hline
  &  relativistic~pressure~radius &   &\\ \noalign{\smallskip}
 5   & d=0.89; r_{0} = 0.00072~{pc} &  1844 & 99.93
\%
\\
\noalign{\smallskip} ~   & f  = 10^{-5}  & ~ & ~
\\
\noalign{\smallskip} ~   & t_{0}=0.249~{yr}; \beta_0= 0.333 & ~ &
~
\\
\noalign{\smallskip}
 \hline
 \hline
 \end{array}
 \]
 \end {table}

\begin{figure*}
\begin{center}
\includegraphics[width=7cm ]{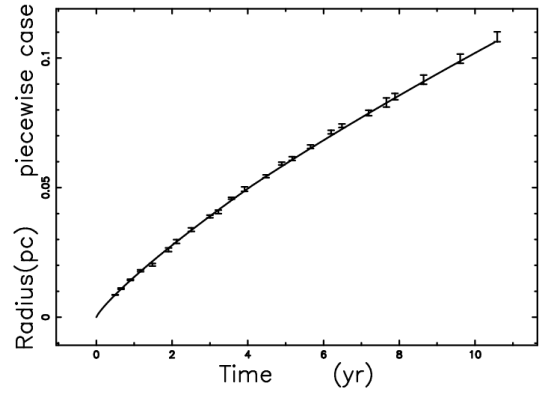}
\end {center}
\caption
{
Theoretical radius as given by the two-regime
fit represented by equation~(\ref{piecewisefit})
with data as in  Table~\ref{datafit} (full line),
and astronomical data of \snr with
vertical error bars.
}
\label{1993break}
    \end{figure*}

A {\em second}  numerical analysis  can be done by assuming
a piecewise function as  in
Fig.  4 of \cite{Marcaide2009}
\begin{equation}
 R(t)   = \left\{ \begin{array}{ll}
             r_{br}(\frac{t}{t_{br}})^{\alpha_1} &
              \mbox {if $t \leq t_{br} $ } \\
             r_{br}(\frac{t}{t_{br}})^{\alpha_2} &
               \mbox {if $t > t_{br} $. }
            \end{array}
            \right.
\label{piecewisefit}
\end{equation}
This type of fit requires the determination
of four parameters, i.e., $t_{br}$, the break time,
$r_{br}$ the radius of expansion at
$t=t_{br}$
and the exponents of the two phases are $alpha_1$
and $alpha_2$.
The two-regime fit can be visualized in
Fig. \ref{1993break}
and Table~\ref{datafit} reports the four parameters.

\begin{figure*}
\begin{center}
\includegraphics[width=7cm ]{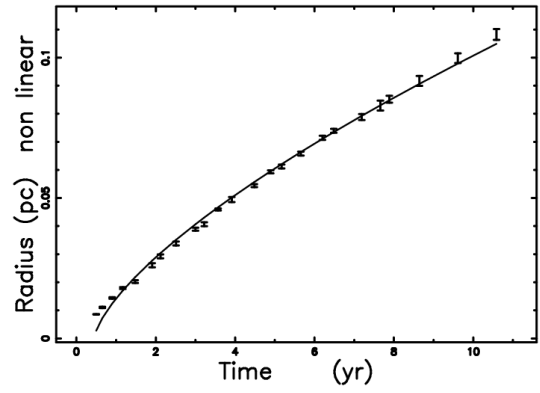}
\end {center}
\caption
{
Theoretical radius as given
by the approximate radius,
equation~(\ref{radiusvarpc}),
with data as in Table~\ref{datafit} (full line).
The  astronomical data of \snr are represented
 with
vertical error bars.
}
\label{1993pc_fit_lev}
    \end{figure*}

A {\em third} type of fit can be done by
adopting the approximate radius as
given by equation  ($\ref{radiusvarpc}$);
Fig.  \ref{1993pc_fit_lev} reports
a fit of this type.

\begin{figure*}
\begin{center}
\includegraphics[width=7cm ]{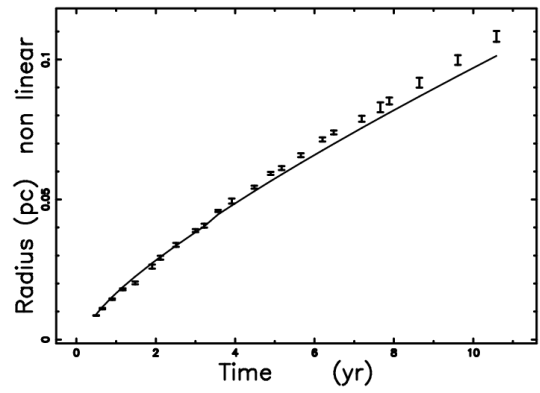}
\end {center}
\caption
{
Theoretical radius as obtained
by the solution of  the nonlinear
equation (\ref{nonlinear})
(full line), data as in Table~\ref{datafit}.
The astronomical data of \snr are represented with
vertical error bars.
}
\label{1993pc_fit_nl}
    \end{figure*}

A {\em fourth } type of fit
implements the nonlinear equation (\ref{nonlinear});
 the four roots
can be found with
the FORTRAN subroutine ZRIDDR from \cite{press};
Fig.  \ref{1993pc_fit_nl} reports
a fit of this type.

\section{Relativistic case}
\label{sec_relativistic}

The relativistic analysis  is split in two.
\begin{enumerate}
\item The relativistic mechanics case
in which the temperature effects are ignored
and the thin layer approximation is used.
\item
The  case of relativistic hydrodynamics in which
the pressure is considered and the velocity is
found from momentum conservation.
\end{enumerate}

\subsection{Relativistic mechanics}

\begin{figure*}
\begin{center}
\includegraphics[width=7cm ]{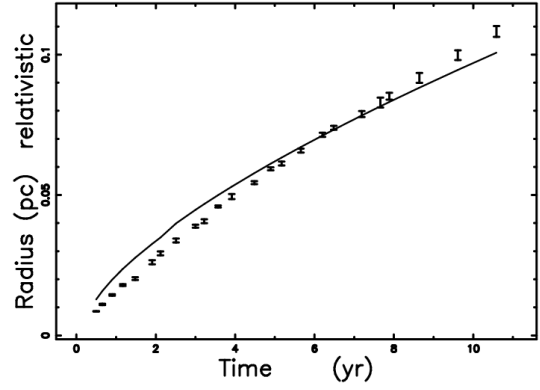}
\end {center}
\caption
{
Theoretical radius as obtained
by the solution of  the relativistic-mechanics
equation (\ref{eqnrel})
(full line), data as in Table~\ref{datafit}.
The astronomical data of \snr are represented
with
vertical error bars.
}
\label{1993pc_fit_rel}
    \end{figure*}

\begin{figure*}
\begin{center}
\includegraphics[width=7cm ]{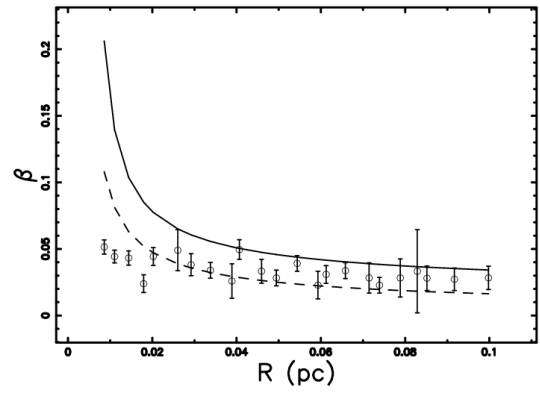}
\end {center}
\caption
{
Theoretical
relativistic-mechanics   velocity
as given by equation   (\ref{velrelativistic}) (dashed line),
theoretical classical velocity
as  given  by equation (\ref{velclassic})      (full  line)
and instantaneous velocity  of \snr  with
uncertainty.
}
\label{1993pc_rel_vel}
    \end{figure*}

Newton's law in special relativity,
after \cite{Einstein1905},
 is:
\begin{equation}
F = \frac {dp} {dt } = \frac {d} {dt} ( m V)
\quad ,
\end{equation}
with
\begin{equation}
 m  = \frac {m_r} {\sqrt{ 1-\frac {v^2}{c_l^2} }}
\quad ,
\end{equation}
where $F$ is the force,
$p$       is the relativistic momentum,
$m$       is the relativistic mass,
$m_r$     is the rest-mass,
$v$       is the velocity
and $c_l$   is the velocity of light,
see  equation~(7.16) in \cite{French1968}.
In the case of the relativistic expansion of a shell
in which all the swept material
resides at two different points,
denoted by radius $ R $  and radius $R_0$,
the previous  equation
gives:
\begin{equation}
M \frac
{ \beta }
{\sqrt  {1-\beta^2}}
=
M_0 \frac
{\beta_0 }
{\sqrt  {1-\beta_0^2}}
\quad ,
\end{equation}
where
$\beta_0$=$v_0/c_l$,
$\beta$=$v/c_l$,
$M$   is the rest mass swept between  0 and $R$ and
$M_0$ is the rest mass swept between  0 and $R_0$.
This  formula is invariant
under Lorentz transformations and the initial
velocity, $v_0$, cannot be greater than the velocity of light.
Assuming a  spatial  dependence of the ISM
as given by formula~(\ref{piecewise}),
relativistic conservation of momentum
gives
\begin{equation}
\frac
{
-4\,\rho\,\pi \, \left( 3\,{r}^{3} \left( {\frac {{\it r_0}}{r}}
 \right) ^{d}-{{\it r_0}}^{3}d \right) \beta
}
{
3\, \left( -3+d \right) \sqrt {1-{\beta}^{2}}
}
=
\frac
{
4\,\rho\,\pi \,{{\it r_0}}^{3}{\it \beta_0}
}
{
3\,\sqrt {1-{{\it \beta_0}}^{2}}
}
\quad .
\end{equation}
According to the previous equation,
$\beta$ is
\begin{eqnarray}
\beta
=
\frac
{
- \left( -3+d \right) {\it \beta_0}\,{{\it r_0}}^{3}
}
{
\sqrt {D_{\beta}}
}
\label{velrelativistic}
\\
with~D_{\beta}=
9\,{R_{{0}}}^{6}{\beta_{{0}}}^{2}-6\,{R_{{0}}}^{6}{\beta_{{0}}}^{2}d
\nonumber \\
 +9 \,{r}^{6-2\,d}{R_{{0}}}^{2\,d}
\nonumber  \\
-6\,{r}^{3-d}{R_{{0}}}^{d+3}d+{R_{{0}}}^ {6}{d}^{2} \nonumber \\
-9\,{\beta_{{0}}}^{2}{r}^{6-2\,d}{R_{{0}}}^{2\,d}+6\,{\beta_
{{0}}}^{2}{r}^{3-d}{R_{{0}}}^{d+3}d \nonumber \quad  .
\end{eqnarray}
In this differential equation of
first order in $r$, the
variables can be separated and the integration
can be expressed as
\begin{equation}
\int_{R_0}^{R} \sqrt {D_{\beta}} dR
=
c \left( 3-d \right) \beta_{{0}}{R_{{0}}}^{3} \left( t-t_{{0}}
 \right)
\quad .
\label{eqnrel}
\end{equation}
The integral of the previous equation
can be performed analytically only in the
cases $d=0$, $d=1$ and $d=2$,
but we do not report the result because
we are interested in a variable value of $d$.
The integral can be easily evaluated
from a theoretical point of view
using the subroutine  QROMB
from \cite{press}.
The numerical result is reported in
Fig. \ref{1993pc_fit_rel}
and the input data in
Table~\ref{datafit}.
The  behavior of relativistic and classical
velocities are reported  in
Fig. \ref{1993pc_rel_vel}.

\subsection{Relativistic hydrodynamics}

\begin{figure*}
\begin{center}
\includegraphics[width=7cm ]{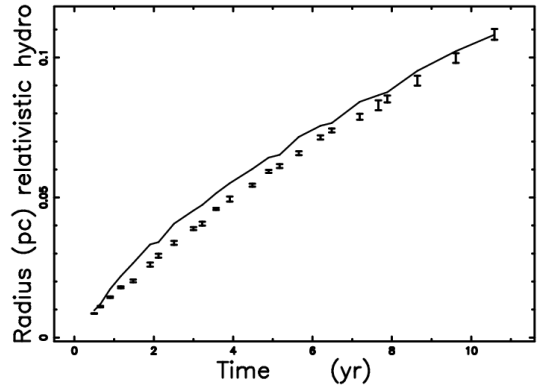}
\end {center}
\caption
{
Theoretical radius as obtained by the solution of  the
 relativistic-hydrodynamics equation (\ref{eqnrelpressure}) (full line)
and
data as in Table~\ref{datafit} with uncertainty.
}
\label{1993pc_fit_pressure}
    \end{figure*}

\begin{figure*}
\begin{center}
\includegraphics[width=7cm ]{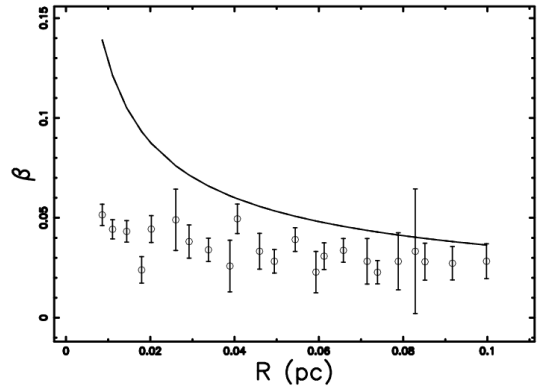}
\end {center}
\caption
{
Theoretical
relativistic-hydrodynamics velocity with pressure
as given by equation   (\ref{betapressure}) (full line),
and instantaneous velocity  of \snr  with
uncertainty.
}
\label{1993pc_rel_vel_pressure}
    \end{figure*}

A relativistic flow on flat space time  is described by the
energy--momentum tensor, $ T^{\mu\nu} $,
\begin{equation}
T^{\mu\nu} = w u^{\mu} u^{\nu} - p g^{\mu\nu} \quad  ,
\end{equation}
where $u^{\mu}$ is the 4-velocity, and the Greek  index  varies
from 0 to 3, $w$ is the enthalpy for unit volume, $p$ is the
pressure and $g^{\mu\nu}$ the inverse metric of the manifold
\cite{Weinberg1972,landau,Hidalgo2005,Gourgoulhon2006}.
Momentum conservation in the presence of velocity, $v$,
along the radial  direction states that
\begin{equation}
(w (\frac{v}{c_l})^2 \frac { 1}{ 1 -\frac {v^2}{c_l^2} } +p) A = cost
\quad , \label{enthalpy}
\end{equation}
where $A(r)$ is the considered surface  area, which is
perpendicular to the motion.
The enthalpy per unit volume is
\begin{equation}
w= c_l^2 \rho  + p \quad ,
\end{equation}
where $\rho$ is the   density, and $c_l$ is
the velocity of light.
The reader  may be puzzled by the $\gamma^2$ factor in
equation~(\ref{enthalpy}), where $\gamma^2= \frac { 1}{ 1 -\frac
{v^2}{c_l^2} }$.
 However it should be remembered  that
$w$ is not an enthalpy, but an enthalpy per unit volume: the extra
$\gamma$ factor arises from  `length contraction' in the direction
of motion \cite{Gourgoulhon2006}. We continue assuming
\begin{equation}
p= \frac{1}{3} f \rho {c_l}^2 \quad , \label{pressure}
\end{equation}
where $f$ is a parameter which has the range $0 \leq f \leq 1$ and is
supposed to be constant during the expansion,
see  formula (2.10.26) in \cite{Weinberg1972}
for a hot extremely
relativistic gas.
The previous
equation (\ref{enthalpy}) becomes
\begin{equation}
 \left( {\frac { \left( {c_l}^{2}\rho+\frac{1}{3}\,f\rho\,
{c_l}^{2} \right) {\beta
}^{2}}{1-{\beta}^{2}}}+\frac{1}{3}\,f\rho\,{c_l}^{2} \right) A =
constant\,.
\end{equation}
The density is  supposed to vary  during the expansion as
\begin{equation}
\rho=\rho_0 (\frac {R_0} {R} )^d
\quad ,
\end{equation}
where $\rho_0$ is the density at $R=R_0$.
 In two
surfaces   of the expansion  we have:
\begin{eqnarray}
- \left( {\frac {R_{{0}}}{R}} \right) ^{d}{R}^{2}{\beta}^{2}
\left( -1 +{\beta}^{2} \right) ^{-1} \nonumber\\
 -1/3\, \left(
{\frac {R_{{0}}}{R}} \right) ^ {d}{R}^{2}f \left( -1+{\beta}^{2}
\right) ^{-1}
\nonumber\\
+{\frac {{R_{{0}}}^{2}{
\beta_{{0}}}^{2}}{-1+{\beta_{{0}}}^{2}}}+1/3\,{\frac
{{R_{{0}}}^{2}f}{ -1+{\beta_{{0}}}^{2}}} \label{conservazionerel}
=0
\quad .
\end {eqnarray}
 The positive solution of the second degree equation is:
\begin{eqnarray}
\beta=\frac{N}{D}
\label{betapressure}
  \\
N= -
(-9\,{R_{{0}}}^{d+2}{R}^{-d+2}{\beta_{{0}}}^{4}+6\,{R_{{0}}}^{2
\,d}{R}^{-2\,d+4}{\beta_{{0}}}^{2}f \nonumber\\
-3\,{R_{{0}}}^{2\,d}{R}^{-2\,d+4}{
\beta_{{0}}}^{4}f \nonumber \\
-6\,{R_{{0}}}^{d+2}{R}^{-d+2}{\beta_{{0}}}^{2}f
 +9\,{R
_{{0}}}^{d+2}{R}^{-d+2}{\beta_{{0}}}^{2} \nonumber \\
-3\,{R_{{0}}}^{2\,d}{R}^{-2\,d
+4}f+3\,{R_{{0}}}^{d+2}{R}^{-d+2}f \nonumber\\
+9\,{R_{{0}}}^{4}{\beta_{{0}}}^{4}+3
\,{R_{{0}}}^{d+2}{\beta_{{0}}}^{4}{R}^{-d+2}f
+6\,{R_{{0}}}^{4}{\beta_{ {0}}}^{2}f \nonumber\\
-{R_{{0}}}^{d+2}{f}^{2}{R}^{-d+2}+{R_{{0}}}^{d+2}{f}^{2}{R}^
{-d+2}{\beta_{{0}}}^{2}+{R_{{0}}}^{4}{f}^{2})^{\frac{1}{2}}
\nonumber\\
D=3\,{R_{{0}}}^{d}{R}^{-d+2}{\beta_{{0}}}^{2} \nonumber\\
-3\,{R_{{0}}}^{d}{R}^{-d+2}
-3\,{R_{{0}}}^{2}{\beta_{{0}}}^{2}-{R_{{0}}}^{2}f \nonumber
 \quad . 
\end{eqnarray}
The equation of motion is
\begin{equation}
\int_{R_0}^{R}  \frac{D}{N}  dR = c \left( t-t_{{0}}
 \right)
\quad . \label{eqnrelpressure}
\end{equation}
The integral is evaluated  using the subroutine QROMB from
\cite{press}. The numerical results are reported in
Figs \ref{1993pc_fit_pressure} and
\ref{1993pc_rel_vel_pressure}.
Fig.  \ref{1993pc_rel_vel_pressure} reports
the decrease of the relativistic velocity.

\section{How the image is formed}

\label{sec_transfer}
In this section, the existing knowledge
about
adiabatic and synchrotron losses,
the acceleration of particles by the Fermi~II mechanism,
, 3D mathematical diffusions with constant
diffusion coefficients
and the rim model with constant density
 are  reviewed
and applied to \snr.
A new  example of a 1D random walk with a step length
equal to the
relativistic electron gyro-radius  is also reported.
A new simple  model for the
temporal evolution of the flux densities is introduced.

\subsection{Acceleration and losses}

\subsubsection{Adiabatic losses}

An ultrarelativistic gas which experiences
an expansion loses energy at the rate
\begin{equation}
-(\frac{dE}{dt}) = \frac{1}{3} ( \nabla \cdot {\bf v} ) E
\quad ,
\end{equation}
where $E$ is the energy and  $ \nabla \cdot {\bf v}  $ is
the divergence of the expansion velocity,
see formula (11.27) in  \cite{longair}.
A simple expression for $\nabla \cdot {\bf v}  $
can be found from the power law model, see
equations (\ref{rpower}) and (\ref{vpower})
\begin{equation}
\nabla \cdot {\bf v} =
\frac
{
{R}^{{\frac {\alpha-1}{\alpha}}}{{\it r_p}}^{\frac{1}{\alpha}} \left( 3\,
\alpha-1 \right)
}
{
R
}
\quad ,
\end{equation}
where $R$ is the temporary radius of the expansion
and   $r_p$ and $\alpha$ are reported in Table \ref{datafit}.
The lifetime, $\tau_{ad}$, of an ultrarelativistic electron
for adiabatic losses is
\begin{equation}
\tau_{ad} = \frac{E} { \frac {dE}{dt}} =
344.39 \, R^{1.22} \, yr
\quad  ,
\end{equation}
when the radius $R$ is  expressed  in pc.
During the ten years of observed expansion,
the radius of \snr  has grown from $\approx$ 0.01 pc
to  $\approx$ 0.1 pc  and therefore the time scale of
adiabatic losses  has increased from
$\approx$ 1.25 yr to  $\approx$ 20.7 yr.

\subsubsection{Synchrotron  losses}

An electron which  loses its  energy  due to
synchrotron radiation
has a lifetime  $\tau_r$, where
\begin{equation}
\tau_r  \approx  \frac{E}{P_r} \approx  500  E^{-1} H^{-2} sec
\quad ,
\label {taur}
\end{equation}
$E$  is the energy in ergs,
$H$ the magnetic field in Gauss,
and  $P_r$  is the total radiated
power, see  formula (1.157) in
\cite{lang}.

\begin{figure*}
\begin{center}
\includegraphics[width=7cm ]{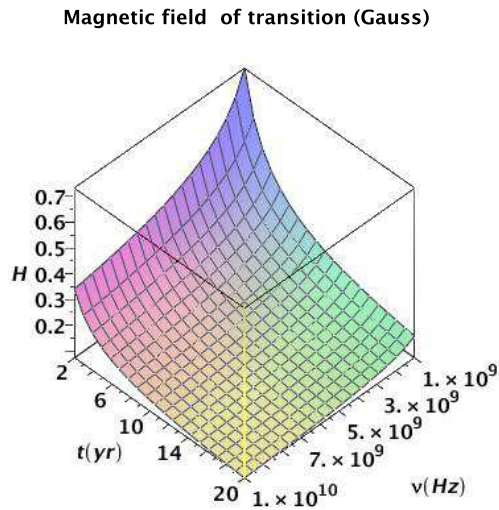}
\end {center}
\caption
{
Magnetic field in Gauss  over which the
synchrotron losses prevail on the adiabatic losses.
}
\label{h_transition}
    \end{figure*}

The  energy  is connected  to  the critical  frequency,
 see formula (1.154)
in \cite{lang},
as
\begin {equation}
\nu_c = 6.266 \times 10^{18} H E^2~Hz
\quad  .
\label {nucritical}
\end{equation}
The lifetime, $\tau_{syn}$,
for synchrotron  losses is
\begin{equation}
\tau_{syn} =
 39660\,{\frac {1}{H\sqrt {H\nu}}} \, yr
\quad  .
\end{equation}
The time-scale of synchrotron losses is  shorter
than that of the adiabatic losses if
the following inequality  is verified
\begin{equation}
\nu >
\frac
{1.572\,10^9  }
{H^3\,\tau_{ad}^2} \quad Hz
\quad ,
\end {equation}
where $H$ is expressed in Gauss and $\tau_{ad}$ in years.
The previous equation can also be expressed as
\begin{equation}
H > 1162.98\,{\frac {1}{\sqrt [3]{{\it \nu}}{{\it t_{ad}}}^{2/3
}}}  Gauss
\quad,
\end{equation}
that  is the magnetic field at which the synchrotron  losses
prevails on the adiabatic
 losses
and Fig. \ref{h_transition}
reports the numerical values of the transition.

\subsubsection{Particle acceleration}

Following \cite{Fermi49,Fermi54},
the gain  in  energy  $\Delta E$
of a particle
which  spirals  around a line of force
is  proportional to its
energy, $E$,
\begin{equation}
\Delta E = B^2 E
\quad  ,
\end{equation}
where $B=u/c_l$ , see  formula (3) in \cite{Fermi49}.
The  continuous form is
\begin  {equation}
\frac {d  E}  {dt }
=
\frac {E }  {\tau }   \quad,
\end {equation}
where $\tau$ is the  typical time-scale.
The probability, $P(t)$ , that the particle
remains in the reservoir
for a period greater than $t$ is now introduced,
\begin {equation}
P(t) = e ^{- \frac {t} {T}}
\quad  ,
\end   {equation}
where $T$  is the time of escape
from the considered region.
The resulting
probability density, $N(E)$,
is
\begin {equation}
N (E)  =
\frac  {\tau} {E_0}
(\frac {E} {E_0})^{-\gamma_f}
\quad  ,
\label {eq:ne}
\end   {equation}
where
$E_0$ is the initial energy
and
\begin {equation}
\gamma_f = 1 + \frac {\tau} {T }
\quad .
\end   {equation}
Equation~(\ref{eq:ne}) can be written as
\begin {equation}
N (E)  =K \;  E ^{-\gamma_f}
\quad  ,
\label {eq:neK}
\end   {equation}
where $K = \frac {\tau} {{E_0}^{-\gamma_f +1}}$~.
A power law spectrum in the particle energy
has now been obtained.
In Fermi~II processes,  the typical time-scale, $\tau_{II}$,
when the particle  stays
in the accelerating region
a time greater than $T$
is
\begin {equation}
\frac{1}{\tau_{II}}  = \frac {4} {3 } ( \frac {u^2} {c_l^2 }) (\frac {c_l } {L })
\quad ,
\label {tau2}
\end   {equation}
where $u$ is the velocity of the accelerating cloud
and $L$  is the mean free path between clouds,
see formula after (4.439) in \cite{lang}.
The mean free path between the accelerating clouds
in the Fermi~II mechanism
can be found from the following inequality:
\begin{equation}
\tau_{II} < \tau_{sync}
\quad  ,
\end{equation}
or
\begin{equation}
L <
1.723 \,10^5 \, \frac{\beta^2} {H \, \sqrt{H \,\nu}}
\quad  pc \quad .
\end{equation}
As an  example, inserting  $\beta=2.82\,10^{-2}$
(value of velocity at $\approx 0.1$\ pc of \snr),
$\nu=1.5 GHz$
and  $H=65.1~Gauss$ (Table 1
in \cite{Marti-Vidal2010})
we obtain
\begin{equation}
L <  6.76 \,10^{-6}
\quad  pc \quad .
\end{equation}
This  model of acceleration can work in the rim model
with constant  density of emitting  particles,
see  Section \ref{secrim}.
In this case the thickness of the emitting region is
$b-a$=0.035 pc which means
that an high number of collisions can be done.
In Fermi~II   process  the
energy  increases exponentially with time
\begin{equation}
E(t) = E_0  \exp (t/\tau)
\quad  ,
\end{equation}
see  , equation (3) in \cite{Fermi54}.
The effect  of the synchrotron losses
on the energy of the electron
during the various  collisions
and the consequent equilibrium  energy
has  been analyzed in \cite{Zaninetti1988b}.
The strong shock accelerating mechanism,
named Fermi~I,
was later
introduced by
\cite{Bell_I,Bell_II}
and produces an
increases in energy  of the
particle of the  type
\begin{equation}
\Delta E   = \frac{u}{c_l} E
\quad ,
\end{equation}
where $u$ is the velocity of the shock,
see  formula (21.20)  in \cite{longair}.
This process allows to predict  a probability
density  in the energy  of  the
accelerated  particles  of  the type  $N(E) \propto E^{-2}$.
In our case  can be considered the process which
accelerates the  particles before they start  to diffuse
from the shock
, see  Section \ref{mathematical} and \ref{secrim}.
A modern review of the two Fermi mechanisms can
be found in \cite{Kulsrud2005,Somov2006,Dermer2009}.

\subsection{The transfer equation}

The transfer equation in the presence of emission only,
see for example
\cite{rybicki}
 or
\cite{Hjellming1988},
 is
 \begin{equation}
\frac {dI_{\nu}}{ds} =  -k_{\nu} \rho I_{\nu}  + j_{\nu} \rho
\label{equazionetrasfer}
\quad ,
\end {equation}
where  $I_{\nu}$ is the specific intensity,
$s$ is the
line of sight, $j_{\nu}$ is the emission coefficient,
$k_{\nu}$ is a mass absorption coefficient,
$\rho$ is the density of mass at position $s$
and the index $\nu$ denotes the frequency of
emission of interest.
The solution to equation~(\ref{equazionetrasfer})
 is
\begin{equation}
 I_{\nu} (\tau_{\nu}) =
\frac {j_{\nu}}{k_{\nu}} ( 1 - e ^{-\tau_{\nu}(s)} )
\quad  ,
\label{eqn_transfer}
\end {equation}
where $\tau_{\nu}$ is the optical depth at frequency $\nu$
\begin{equation}
d \tau_{\nu} = k_{\nu} \rho ds
\quad.
\end {equation}

The volume emissivity (power
per unit frequency interval
per unit volume per unit solid angle)
of the ultrarelativistic radiation
from a group of electrons,
according to
\cite{lang},
 is
\begin {equation}
\epsilon (\nu) =
\int  P(\nu) N(E) dE
\quad ,
\end{equation}
where $P(\nu)$ is the total power radiated
per unit frequency interval by one electron
and  $N(E) dE$ is the number of electrons
per unit  volume,
per unit solid angle along the line
of sight that are moving in the direction
of the observer and whose energies lie in the range
$E$ to $E+dE$.
In the case of a power law  spectrum,
\begin{equation}
N(E)dE = K E^{-\gamma_f}
\label{spectrum}
\quad  ,
\end{equation}
where $K$ is a constant.
The value  of the constant  $K$ can be found
by assuming that
the  probability density function
for the relativistic energy
is of  Pareto type
as defined in   \cite{evans}
\begin {equation}
f(x;a,c_p) = {c_p a^{c_p}}{x^{-(c_p+1)}} \quad ,
\label{pareto}
\end {equation}
with $ c_p~> 0$.
In our case,
$c_p=\gamma_f-1$ and  $a=E_{min}$,
where $E_{min}$ is the minimum energy.
We can now extract
\begin{equation}
K=
N_0 (\gamma_f -1) E_{min}^{\gamma_f -1}
\quad  ,
\end{equation}
where $N_0$ is the total number
of relativistic electrons  per unit volume,
here assumed to be approximately
equal to the matter number density.
The previous  formula can also be expressed
as
\begin{equation}
K=
\frac {\rho}{1.4 \, m_{\mathrm {H}}}
 (\gamma_f -1) E_{min}^{\gamma_f -1}
\quad  ,
\end{equation}
where $m_H$ is the mass of hydrogen.
The emissivity  of  the ultrarelativistic
synchrotron radiation from a homogeneous
and isotropic distribution of electrons
whose  $N(E)$  is given by
equation~(\ref{spectrum})
is, according to
\cite{lang},
\begin{eqnarray}
j_{\nu} \rho  =  \\
\approx 0.933 \times 10^{-23}
\alpha (\gamma_f) K H_{\perp} ^{(\gamma_f +1)/2 }
\times \nonumber \\ \times \bigl (
 \frac{6.26 \times 10^{18} }{\nu}
\bigr )^{(\gamma_f -1)/2 } \nonumber    \\
erg\, sec^{-1} cm^{-3} Hz^{-1} rad^{-2}
\nonumber
\quad ,
\end{eqnarray}
where $\nu$ is the frequency
and   $\alpha (\gamma_f)$  is a slowly
varying function
of $\gamma_f$ which is of the order of unity
and is given by
\begin{eqnarray}
\alpha(\gamma_f) =  \nonumber \\
2^{(\gamma_f -3)/2} \frac{\gamma_f+7/3}{\gamma_f +1}
\Gamma \bigl ( \frac {3\gamma_f -1 }{12} \bigr )
\Gamma \bigl ( \frac {3\gamma_f +7 }{12} \bigr )
\quad ,
\end{eqnarray}
for  $\gamma_f \ge \frac{1}{2}$.

We now continue to analyze the case of an optically thin layer
in which $\tau_{\nu}$ is very small
(or $k_{\nu}$ is very small)
and where the the density $\rho$ is substituted
for the concentration $C(s)$
 of relativistic electrons
\begin{equation}
j_{\nu} \rho =K_e  C(s)
\quad  ,
\end{equation}
where $K_e$ is a constant function
of the energy power law index,
magnetic field
and frequency of e.m.\ emission.
The intensity is now
\begin{equation}
 I_{\nu} (s) = K_e
\int_{s_0}^s   C (s\prime) ds\prime \quad  \mbox {Optically thin layer}
\quad.
\label{transport}
\end {equation}
The increase in brightness
is proportional to the concentration
integrated along
the line of  sight.
In  numerical  experiments,
the concentration is memorized
on
the  visitation-grid
${\mathcal S}$ and the intensity is
\begin{eqnarray}
{\it I}\/(i,j) = \sum_k  \triangle\,s \times  {\mathcal S}(i,j,k)
\label{thin}
\\
\quad  \mbox {Optically thin layer}\quad,
\nonumber 
\end{eqnarray}
where $\triangle$s is the spatial interval between
the various values and  the sum is performed
over the   interval of existence of index $k$.
The theoretical flux density is then obtained by integrating
the intensity at a given frequency  over the solid angle of the
source.
In order to deal with the transition to
the optically thick case,
the intensity is given
by
\begin{eqnarray}
\label{transition} {\it I}\/(i,j) = 
\frac {1}{K_a}  (1 - \exp (-
K_a\sum_k
\triangle\,s \times {\mathcal S}(i,j,k)))  \\
 \quad
\mbox {Thin $\longmapsto $ Thick }\quad, \nonumber
\end{eqnarray}
where  $K_a$ is a constant that represents the absorption.
Considering the  Taylor expansion of the last
formula (\ref{transition}),
equation~(\ref{thin}) is obtained.

\subsection{3D diffusion from a spherical source}
\label{mathematical}

Once the concentration, $C$, and diffusion
coefficient, $D$,
are introduced, Fick'~s law
in three dimensions  is
\begin{equation}
\frac {\partial C }{\partial t} =
D \nabla^2 C
\quad.
\label{eqfick}
\end {equation}
Under steady-state conditions,
\begin{equation}
D \nabla^2 C   = 0
\quad .
\label{eqfick_steady}
\end {equation}

\begin{figure*}
\begin{center}
\includegraphics[width=7cm ]{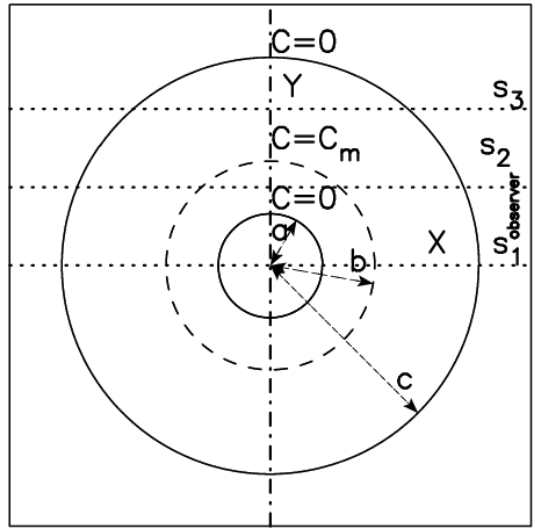}
\end {center}
\caption
{
The spherical source
is  represented
by
the dashed line and the two absorbing boundaries
by full lines.
The observer is situated along the $x$ direction, and
three lines of sight are indicated.
Adapted from Fig. 3.1 in
Berg (1993).
}
\label{plot}
    \end{figure*}

The  concentration rises from 0 at {\it r=a}  to a
maximum value $C_m$ at {\it r=b} and then  falls again
to 0 at {\it  r=c}.
The  solution of  equation~(\ref{eqfick_steady})
 is
\begin{equation}
C(r) = A +\frac {B}{r}
\quad,
\label{solution}
\end {equation}
where $A$ and $B$  are determined
by the boundary conditions,
\begin{equation}
C_{ab}(r) =
C_{{m}} \left( 1-{\frac {a}{r}} \right)  \left( 1-{\frac {a}{b}}
 \right) ^{-1}
\quad a \leq r \leq b
\quad,
\label{cab}
\end{equation}
and
\begin{equation}
C_{bc}(r)=
C_{{m}} \left( {\frac {c}{r}}-1 \right)  \left( {\frac {c}{b}}-1
 \right) ^{-1}
\quad b \leq r \leq c
\quad.
\label{cbc}
\end{equation}
These solutions can be found in
\cite{berg}
or in
\cite{crank}.
Fig. \ref{plot} shows
a spherical shell source of radius  $b$
between a spherical absorber
of radius $a$ and a spherical
absorber of radius $c$.
The  concentration rises from 0 at {\it r=a}  to a
maximum value $C_m$ at {\it r=b} and then  falls again
to 0 at {\it  r=c}.
The concentrations to be used
are formulas (\ref{cab}) and  (\ref{cbc})  once
$r=\sqrt{x^2+y^2}$ is imposed;
these two concentrations are  inserted in
formula~(\ref{eqn_transfer})  which  represents
the transfer equation.
 The geometry of the phenomenon fixes
 three different zones ($0-a,a-b,b-c$) for the variable $y$,
 see~\cite{Zaninetti2007_c};
 the first segment, $I^I(y)$, is
\begin{eqnarray}
I^I(y)=  
\nonumber  \\ 
2 {\frac {b{\it C_m} \sqrt {{a}^{2}-{y}^{2}}}{-b+a}}-2 {\frac {b{
\it C_m} a\ln  \left( \sqrt {{a}^{2}-{y}^{2}}+a \right) }{-b+a}}
\nonumber \\
-2 {
\frac {b{\it C_m} \sqrt {{b}^{2}-{y}^{2}}}{-b+a}}+2 {\frac {b{\it C_m}
 a\ln  \left( \sqrt {{b}^{2}-{y}^{2}}+b \right) }{-b+a}}
\nonumber \\
+2 {\frac {b
{\it C_m} c\ln  \left( \sqrt {{b}^{2}-{y}^{2}}+b \right) }{-c+b}}-2 {
\frac {b{\it C_m} \sqrt {{b}^{2}-{y}^{2}}}{-c+b}}
\nonumber \\
-2 {\frac {b{\it C_m}
 c\ln  \left( \sqrt {{c}^{2}-{y}^{2}}+c \right) }{-c+b}}+2 {\frac {b
{\it C_m} \sqrt {{c}^{2}-{y}^{2}}}{-c+b}}
\label{I_1} \\
~ 0 \leq y < a \quad.  \nonumber
\end{eqnarray}

The second segment, $I^{II}(y)$, is
 \begin{eqnarray}
 I^{II}(y)=  
-{\frac {b{\it C_m} a\ln  \left( {y}^{2} \right) }{-b+a}}-2 {\frac {b
{\it C_m} \sqrt {{b}^{2}-{y}^{2}}}{-b+a}}
\nonumber \\
+2 {\frac {b{\it C_m} a\ln 
 \left( \sqrt {{b}^{2}-{y}^{2}}+b \right) }{-b+a}}
\nonumber \\
+2 {\frac {b{\it C_m
} c\ln  \left( \sqrt {{b}^{2}-{y}^{2}}+b \right) }{-c+b}}
\nonumber \\
-2 {\frac {
b{\it C_m} \sqrt {{b}^{2}-{y}^{2}}}{-c+b}}-2 {\frac {b{\it C_m} c\ln 
 \left( \sqrt {{c}^{2}-{y}^{2}}+c \right) }{-c+b}}
\nonumber  \\
+2 {\frac {b{\it C_m
} \sqrt {{c}^{2}-{y}^{2}}}{-c+b}}
\label{I_2} \\
  a \leq y < b  \quad. \nonumber 
\end{eqnarray}
The third segment, $I^{III}(y)$, is
 \begin{eqnarray}
 I^{III}(y)=   
 \nonumber  \\
{\frac {b{\it C_m} c\ln  \left( {y}^{2} \right) }{-c+b}}-2 {\frac {b{
\it C_m} c\ln  \left( \sqrt {{c}^{2}-{y}^{2}}+c \right) }{-c+b}}
\nonumber \\
+2 {
\frac {b{\it C_m} \sqrt {{c}^{2}-{y}^{2}}}{-c+b}}
\nonumber \\
 b \leq y < c  \quad.  
\label{I_3}
\end{eqnarray}

\begin{figure*}
\begin{center}
\includegraphics[width=7cm ]{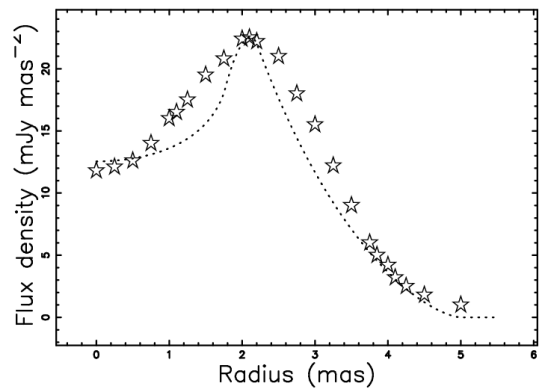}
\end {center}
\caption
{
 Cross-section of the mathematical  intensity ${\it I}$
 (formulas (\ref{I_1}), (\ref{I_2}) and  (\ref{I_3})),
 through the center  (dotted  line) of \snr
 and  real data  (empty stars),
 $\chi^2$ = 100.49
The real data made on  day  1889 after  the explosion
have been extracted  by the author
from Fig.  3 of
Marcaide et al.   (2009).
Parameters as in Table~\ref{dataabc}.
}
\label{pn_cut}
    \end{figure*}

\begin{figure*}
\begin{center}
\includegraphics[width=7cm ]{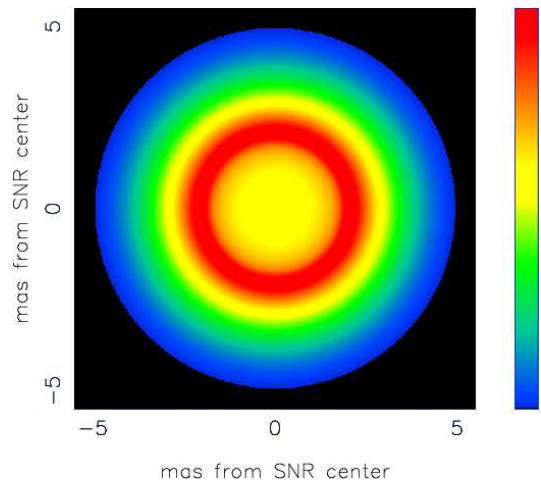}
\end {center}
\caption
{
Contour map  of  ${\it I}$
adjusted to simulate  \snr.
Parameters as in Table~\ref{dataabc}.
}
\label{pnbri}
    \end{figure*}

The profile of ${\it I}$
made up of the three
segments (\ref{I_1}), (\ref{I_2}) and  (\ref{I_3}),
can be calibrated against the real data
of \snr  and an acceptable
match can be achieved by adopting the parameters
reported in Table~\ref{dataabc}.
 \begin{table}
 \caption[]{Simulation of \snr by 3D diffusion,
  optically thin case}
 \label{dataabc}
 \[
 \begin{array}{lll}
 \hline
 \hline
 \noalign{\smallskip}
 symbol  & meaning & value  \\
 \noalign{\smallskip}
 \hline
 \noalign{\smallskip}
a  &  radius ~internal~sphere    & 1.76 (mas)  \\ \noalign{\smallskip}
b  & radius~shock                        & 2.2  (mas)  \\ \noalign{\smallskip}
c  & radius~external~sphere    & 5.0  (mas)  \\ \noalign{\smallskip}
\frac {I_{limb}} {I_{center}} &  ratio~observed~ intensities  &
1.7926       \\ \noalign{\smallskip}
\frac {I_{max}} {I(y=0)} & ratio~ theoretical~ intensities  &
1.7927       \\ \noalign{\smallskip}
 \hline
 \hline
 \end{array}
 \]
 \end {table}
The theoretical intensity can therefore
be plotted as a function of the distance from
the center, see Fig. \ref{pn_cut},
or as  a contour map,
see Fig. \ref{pnbri}.

\begin{figure*}
\begin{center}
\includegraphics[width=7cm ]{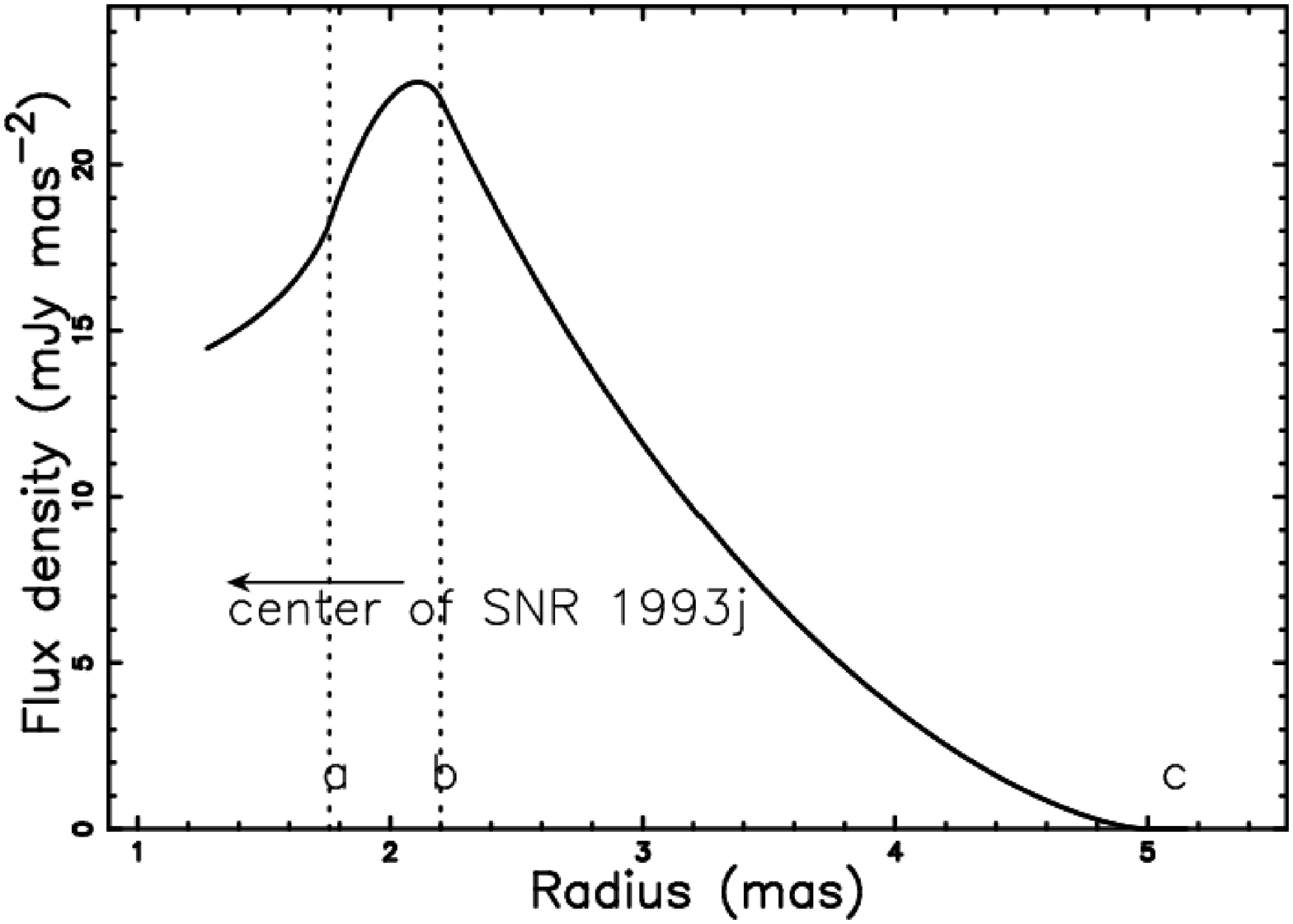}
\end {center}
\caption
{
 Cross-section through the mathematical  intensity ${\it I}$
 towards the edge of \snr.
 The three  parameters which  characterize
 the expanding PN, {\it a  }, {\it b} and  {\it c},
 are reported.
Parameters as in Table~\ref{dataabc}.
}
\label{pn_cut_exp}
    \end{figure*}

The position of  the minimum
of ${\it I}$  is at $y=0$ and the position of the maximum
is situated in the region $a \leq y < b $,
or more precisely at:
\begin{equation}
y={\frac {\sqrt {- \left( b-2\,a+c \right) a \left( ab-2\,bc+ac
\right) }}{b-2\,a+c}}
\quad.
\label{maximum}
\end{equation}
This  means that the maximum emission  is not at the
position of the shock, identified here as $b$,
but shifted a little towards the center;
see Fig. \ref{pn_cut_exp}.

The ratio between the theoretical maximum intensity, $I_{max}$,
as given by formula~(\ref{maximum}), and minimum intensity ($y=0$)
is complex and is reported
in formulas (74-76) in 
\cite{Zaninetti2007_c}.
The observed ratio as well as
the theoretical ratio are
reported in Table~\ref{dataabc}.

\begin{figure*}
\begin{center}
\includegraphics[width=7cm ]{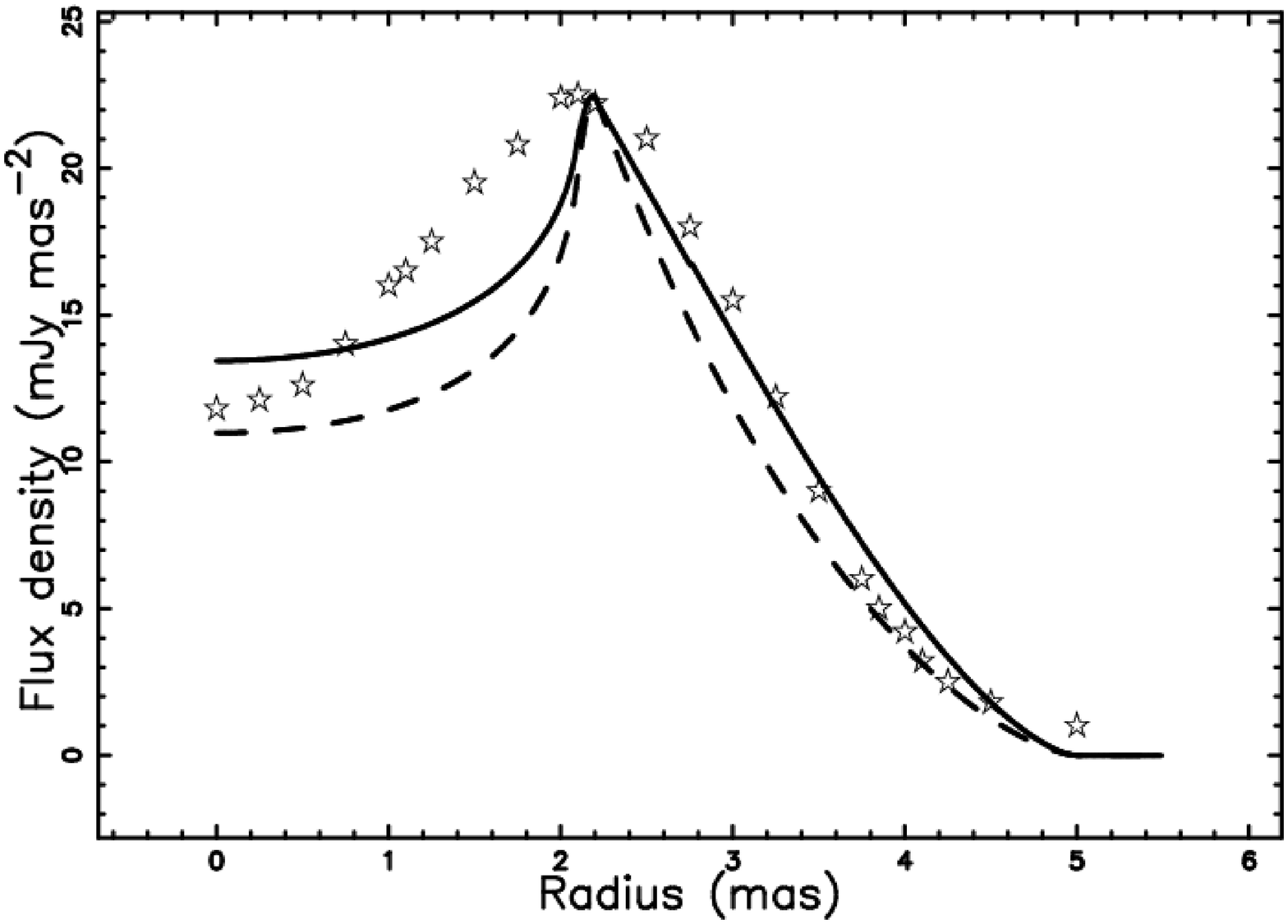}
\end {center}
\caption
{
 Cross-section through the mathematical intensity ${\it I}$
 (formulas (\ref{I_1}), (\ref{I_2}) and  (\ref{I_3})),
in the optically thin case
(dashed line, $\chi^2=237.3$),
and optically thick case (full line,
$\chi^2=84.7$)
 and  real data (empty stars).
Parameters as in Table~\ref{dataabcabsorption}.
}
\label{absorption}
    \end{figure*}

The effect of  absorption is easily evaluated
by applying  formula~(\ref{transition})  and fixing
the value of $K_a$.
The result  is
shown
in Fig. \ref{absorption}.

 \begin{table}
 \caption[]{Simulation of \snr with 3D diffusion,
  optically thick case with $K_a=0.2$.}
 \label{dataabcabsorption}
 \[
 \begin{array}{llc}
 \hline
 \hline
 \noalign{\smallskip}
 symbol  & meaning & value  \\
 \noalign{\smallskip}
 \hline
 \noalign{\smallskip}
a  & radius~internal~sphere    & 2.01 (mas)  \\ \noalign{\smallskip}
b  & radius~of~shock                        & 2.2  (mas)  \\ \noalign{\smallskip}
c  & radius~external~sphere    & 5.0  (mas)  \\
\noalign{\smallskip}
\frac {I_{limb}} {I_{center}} &  ratio~observed~ intensities  &
1.7926       \\ \noalign{\smallskip}
\frac {I_{max}} {I(y=0)} & ratio~optically~ thin~case &
2.0491      \\ \noalign{\smallskip}
\frac {I_{max}} {I(y=0)} & ratio~optically~ thick~case&
1.6741      \\ \noalign{\smallskip}
 \hline
 \hline
 \end{array}
 \]
 \end {table}

\subsection{The rim model with constant density}

\label{secrim}

\begin{figure*}
\begin{center}
\includegraphics[width=7cm ]{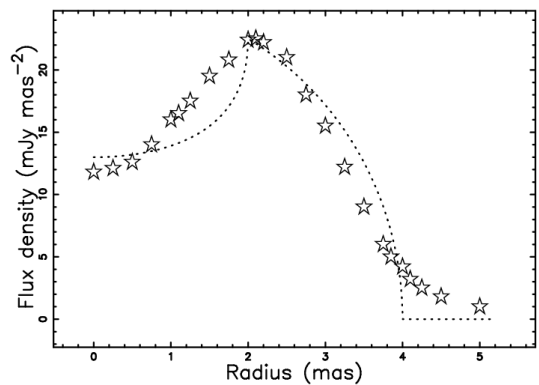}
\end {center}
\caption
{
 Cut of the mathematical  intensity ${\it I}$
 of the rim  model ( equation~(\ref{irim}))
 through the center  (dotted  line) of \snr
 and  real data  (empty stars).
 The parameters  are
 $a=2  $ mas ,  $b=4 $ mas
, $\frac {I(y=a)} {I(y=0)}$=1.73
and   $\chi^2$ = 125.19.
}
\label{ring_cut}
    \end{figure*}

We assume that the number density
of ultrarelativistic electrons
$C$ is constant and in particular
rises from 0 at $r=a$ to a maximum value $C_m$ , remains
constant up to $r=b$ and then falls again to 0,
see Section 5.2 in \cite{Zaninetti2009a}.
The length of sight , when the observer is situated
at the infinity of the $x$-axis ,
is the locus
parallel to the $x$-axis which  crosses  the position $y$ in a
Cartesian $x-y$ plane and terminates at the external circle
of radius $b$.
The locus length is
\begin{eqnarray}
l_{0a} = 2 \times ( \sqrt { b^2 -y^2} - \sqrt {a^2 -y^2})
\quad  ;   0 \leq y < a  \nonumber  \\
l_{ab} = 2 \times ( \sqrt { b^2 -y^2})
 \quad  ;  a \leq y < b    \quad .
\label{length}
\end{eqnarray}
When the number density
of ultrarelativistic electrons
 $C_m$ is constant between two spheres
of radius $a$ and $b$
the intensity of radiation is
\begin{eqnarray}
I_{0a} =
\nonumber   \\
C_m \times 2 \times ( \sqrt { b^2 -y^2} - \sqrt {a^2 -y^2})
\quad  ;   0 \leq y < a    \\
I_{ab} =C_m \times  2 \times ( \sqrt { b^2 -y^2})
 \quad  ;  a \leq y < b    \quad .
\label{irim}
\end{eqnarray}
The ratio between the theoretical intensity at the maximum , $(y=a)$ ,
 and at the minimum , ($y=0$) ,
is given by
\begin{equation}
\frac {I(y=a)} {I(y=0)} = \frac {\sqrt {b^2 -a^2}} {b-a}
\quad .
\label{ratioteorrim}
\end{equation}
The  parameter $b$ is identified with
the external radius
of the SNR.
The parameter $a$ can be found from
the following formula
\begin{equation}
a  = \frac
{
b \left(   (\frac {I(y=a)} {I(y=0)})_{obs}^2 - 1  \right)
}
{
\left(   (\frac {I(y=a)} {I(y=0)})_{obs}^2 + 1  \right) }
\quad ,
\end{equation}
where  $(\frac {I(y=a)} {I(y=0)})_{obs} $
is the observed ratio between maximum
intensity at  the rim
and intensity at the center.
A cut in the theoretical intensity
is reported in Figure~\ref{ring_cut}.

\subsection{1D diffusion}

The Fick equation in 1D with a constant diffusion
coefficient, see \cite{crank}, is
\begin{equation}
\frac {d^2C}{dr^2} =0
\quad .
\label{fick_1D}
\end {equation}
The general solution to
equation~(\ref{fick_1D})
is
\begin{equation}
C(r) = A + B r
\quad .
\label{solution_1D}
\end{equation}
The boundary conditions  give
\begin{equation}
C(r) =
C_{{m}}  \frac {r-a}{b-a}
\quad a \leq r \leq b
\quad ,
\label{cab_1d}
\end{equation}
and
\begin{equation}
C(r) =
C_{{m}}   \frac {r-c}{b-c}
\quad b \leq r \leq c
\quad .
\label{cbc_1d}
\end{equation}
The transport of relativistic electrons
with a step length
equal to the gyro-radius of the relativistic electrons
is called Bohm diffusion (\cite{Bohm})
and the diffusion coefficient is energy-dependent.
The assumption of Bohm diffusion
allows of setting a one-to-one
correspondence between the energy and the step length
in the random walk.
The relativistic electron  gyro-radius $r_H$ is
\begin{equation}
r_H =\frac {m_e c_l v_{\perp} \gamma_e} {qB}
\quad ,
\end{equation}
where $m_e$  is the electron mass,
$gamma_e$ is the Lorentz  factor of the relativistic electron,
$c_l $ is the velocity of light,
$v_{perp}$ is the velocity perpendicular to the magnetic field,
$B$ is the magnetic field in Gauss and
$q$ is the electron charge in statcoulombs.
The astrophysical  version, see formula 1.153
in \cite{lang}, is
\begin{equation}
r_H = 2 \, 10^9 \frac{E}{B}  cm
\quad ,
\end{equation}
where $E$ is the electron energy in cgs.
The typical frequency of synchrotron emission,
$\nu_c$, is
\begin{equation}
\nu_c \approx  6.266 \, 10^{18} B E^2  Hz
\quad .
\end{equation}
This formula  allows us to express the
relativistic electron  gyro-radius  as a function
of the wavelength of emission expressed
in centimeters, $\lambda_1$,
\begin{equation}
r_H  =  \frac {1.42 \,10^{-6}\,\sqrt {{\frac {{\it B_{-5}}}{{\it \lambda_1}}}}}
              {{{\it B_{-5}}}^{2}} pc
\quad ,
\end{equation}
where
the magnetic
field,
$H_{-5}$,
is
expressed in units of $10^{-5}$ Gauss.

\begin{figure}
\includegraphics[width=7cm ,angle=0]{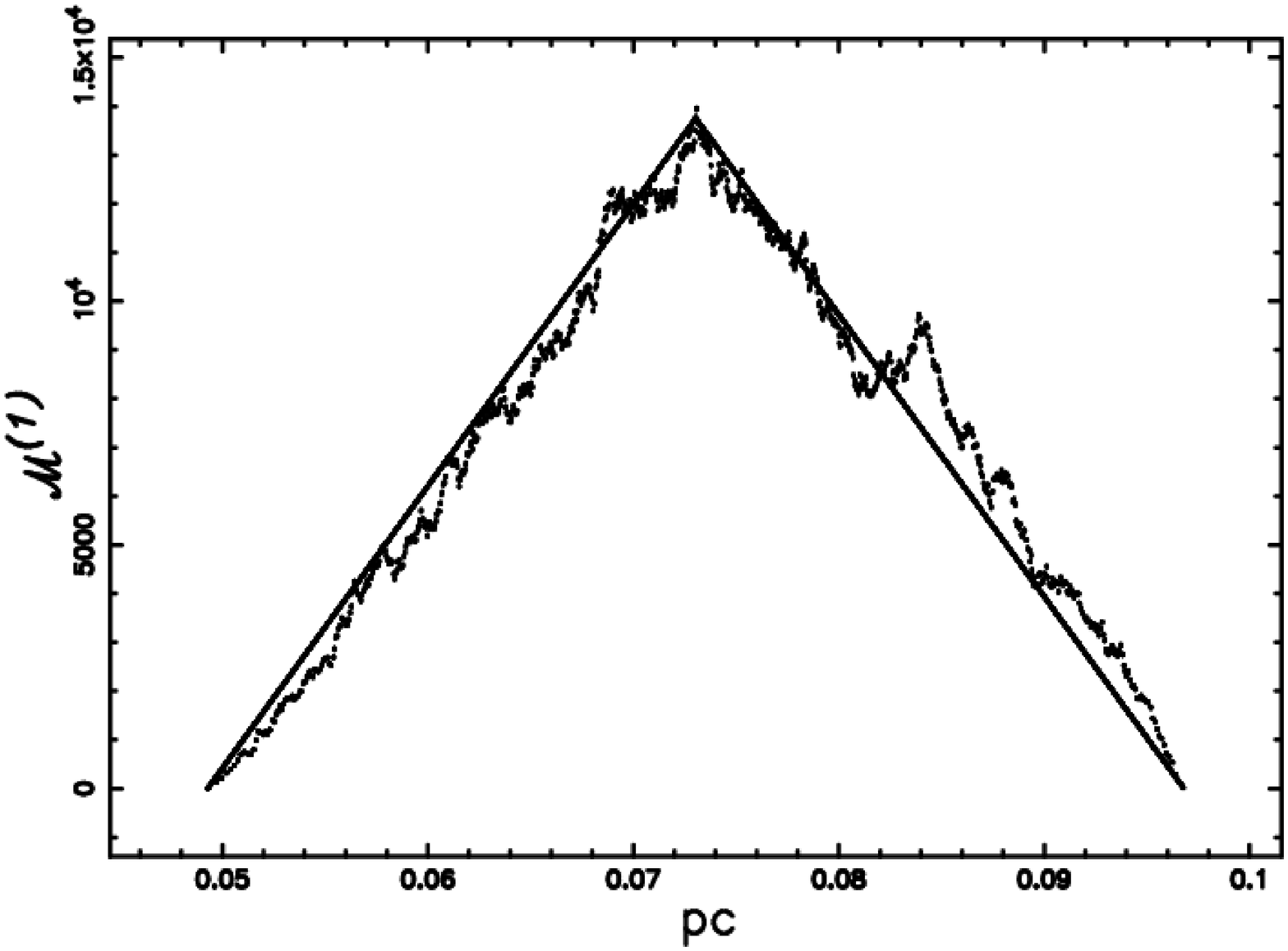}
\caption
{
Values of  concentration  computed with
equation~(\ref{cbc_1d}) (full line) compared with the results
of a Monte Carlo simulation
(filled circles).
The parameters in the  Monte Carlo 1D random
walk  are
$H_5$ = 0.5, $ \lambda_1$ = 6,
$r_H = 1.65 10^{-6}$ pc
and NTRIALS = 100.
}
\label{soluz_1d}
    \end{figure}

\begin{figure*}
\begin{center}
\includegraphics[width=7cm ]{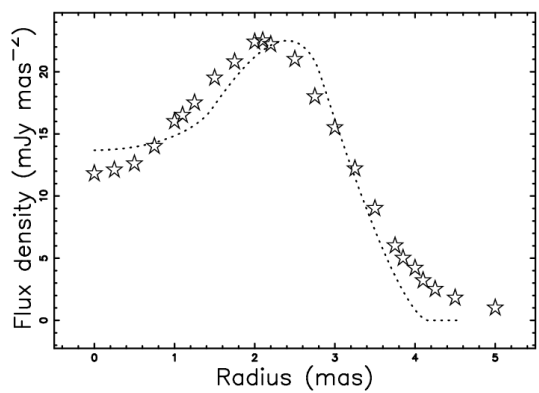}
\end {center}
\caption
{
 Cross-section through the mathematical
 intensity ${\it I}$ of
 1D diffusion
 in the optically thin case (dashed line,
 $\chi^2=80.64$)
 and  real data    (empty stars).
Parameters as in Table~\ref{dataabccut_triangolo}.
}
\label{cut_triangolo}
    \end{figure*}

The 1D theoretical solution and
a Monte Carlo
simulation characterized by a given number
of trials, NTRIALS,
are reported in Fig. \ref{soluz_1d}.
The theoretical intensity can therefore be
evaluated by a numerical integration
of equations (\ref{cab_1d}) and  (\ref{cbc_1d})
along the line of sight,
see Fig. \ref{cut_triangolo}
and Table \ref{dataabccut_triangolo}.

 \begin{table}
 \caption[]{Simulation of \snr based on 1D diffusion,
  optically thin case. }
 \label{dataabccut_triangolo}
 \[
 \begin{array}{llc}
 \hline
 \hline
 \noalign{\smallskip}
 symbol  & meaning & value  \\
 \noalign{\smallskip}
 \hline
 \noalign{\smallskip}
a  & radius~internal~sphere    & 1.44 (mas)  \\ \noalign{\smallskip}
b  & radius~of~shock                        & 2.79 (mas)  \\ \noalign{\smallskip}
c  & radius~external~sphere    & 4.15 (mas)  \\
\noalign{\smallskip}
\frac {I_{limb}} {I_{center}} &  ratio~observed~ intensities  &
1.7926       \\ \noalign{\smallskip}
\frac {I_{max}} {I(y=0)} & ratio~optically~ thin &
1.6430      \\ \noalign{\smallskip}
 \hline
 \hline
 \end{array}
 \]
 \end {table}

\subsection{Evolution of flux densities}

\begin{figure*}
\begin{center}
\includegraphics[width=7cm]{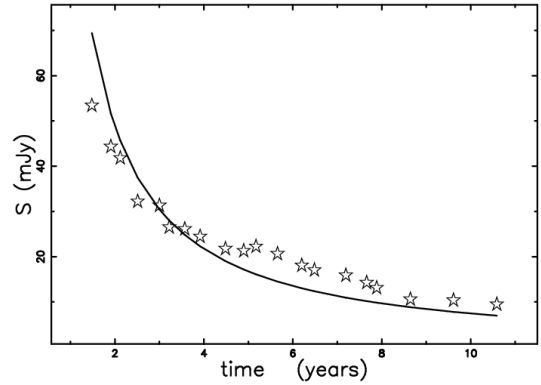}
\end {center}
\caption
{
 Observed time evolution of total
 flux densities of  \snr
 as  given by  VLBI data (empty stars) at 6 cm
 and  theoretical relationship as
 given by formula (\ref{fluxtime}) (full line),
 with  $d$ = 2.79  and  $\alpha_p = 0.82$.
 The VLBI data at 6 cm are extracted from Table 1
 in Marcaide  et al.\ 2009.
}
\label{flux}
    \end{figure*}

The source of synchrotron luminosity
is assumed here to be the flux of kinetic energy,
$L_m$,
\begin{equation}
L_m = \frac{1}{2}\rho 4 \pi R^2 V^3
\quad ,
\end{equation}
where $R$  is the instantaneous radius of the SNR and
$\rho$  is the density in the advancing layer
in which synchrotron emission takes place.
The density in the advancing  layer is  assumed
to scale as  $R^{-d}$, see formula (\ref{piecewise}),
which means that
\begin{equation}
L_m  \propto R ^{2-d}  V^3
\quad .
\end{equation}
The temporal and velocity evolutions
can be
given by the  power law
dependencies of equations (\ref{rpower}) and
(\ref{vpower})
and therefore
\begin{equation}
L_m  \propto  t^{-\alpha_p d+5\alpha_p-3}
\quad  .
\end{equation}
The  synchrotron luminosity
$L_{\lambda}$
and  the observed  flux  $S_{\lambda}$
at a given wavelength    $\lambda$
are assumed   to be proportional
to the  mechanical  luminosity
and therefore
\begin{equation}
S_{\lambda} =  S_0 (\frac {t}{t_0})^{-\alpha_p d+5\alpha_p-3}
\quad ,
\label{fluxtime}
\end{equation}
where $S_0$ is the  flux at  $t=t_0$.
The availability of synchrotron  flux
at  6~cm, see Table 1 and Fig.  14
in \cite{Marcaide2009}, allows of fixing the parameters $S_0$.
Fig.  \ref{flux} reports  the
observed flux  as a function of time
as well as a theoretical evaluation
using equation (\ref{fluxtime}).
The astrophysical version of the above equation
is
\begin{equation}
S_{\lambda=6cm} =  69.41 \,{(\frac{t}{t_0}})^{- 1.16}        ~  mJy
\quad ,
\label{fluxtimeastro}
\end{equation}
with the time expressed in years.

\section{Conclusions}

{\bf Law of motion}
The  first two parts  of the law of motion
for SNR are  thought to be a free expansion
in which $R \propto  t$ and an energy  conserving
phase in which   $R \propto  t^{2/5}$.
A careful analysis of \snr  in the first 10 $yr$
conversely  suggest that $R \propto  t^{0.82}$.
In other words  the  free expansion  which  follows
the first Newton's law of motion does not corresponds
to the observations.
This  observational evidence requires a law of motion
which  contains an adjustable  parameter.
Here we have chosen as a physical arguments
the thin layer approximation in a medium which has
a  decreasing density
of the type $\propto  R^{-d}$
The two laws of motion  here deduced
are the classical  nonlinear equation (\ref{nonlinear}) and
the complex relativistic-mechanics  equation (\ref{eqnrel}).
In the classical case of the thin layer approximation,
we derived
a useful  asymptotic law
of the type $R \propto t^{1/(4-d)} $, see equation
\ref{asymptotic}.
This means that $d=1-1/(\alpha)$ and therefore
the value of the $d$  can be deduced from the
observational parameter $\alpha$.
Both the classical  and relativistic equations
of the thin layer approximation
can
be used as a fitting function
to deduce  the remaining  physical parameters ,
which are
the initial radius and  the
initial velocity or $\beta$.

A third law of motion is deduced in the framework
of the relativistic-hydrodynamics with pressure
applying the momentum conservation,
see equation (\ref{eqnrelpressure}).
The  two relativistic cases
here considered does  not have an asymptotic  solution.
The evaluation of the merit function $\chi^2$
, see equation (\ref{chisquare}) as well
the efficiency of the final radius,
see equation (\ref{efficiency})
allows to eliminate the free expansion
+ the Sedov phase which are supposed  to characterize
the first two phases  of the SNR's.
The best results are obtained
by the "ad hoc" piecewise   function (\ref{piecewisefit})
followed
by the classical  nonlinear equation   (\ref{nonlinear})
once the  $\chi^2$ and efficiency
are considered together.

{\bf Relativistic velocities}
It is  really necessary the relativistic treatment
for the equation of motion?
We  briefly recall that at the moment of writing
the maximum observed velocity
from
CaII H\&K absorptions  in  \sn2002bo
is  $\approx 26000 km/s$ as  measured
at  -15 days from B maximum ,
see Figure 11  in \cite{Benetti2004}.
A  shift of 7 days  characterizes   the time delay between
Gamma-ray burst  (GRB) and SN spectrum
in the case of
 \s2003dh , see
\cite{Matheson2003}.
A quadratic fit  allows
to extrapolate  at t=-22 days from B maximum
the  velocity of expansion
; in particular  we found
$\approx 28710 km/s$   as a  maximum velocity
for \sn2002bo
,
see data in Figure 11b of \cite{Benetti2004}.
This means that $\beta\approx 1/10.4$
which is not far from the canonical $\beta =1/10 $
which marks the transition from classical
to relativistic regime.

{\bf Formation of the image}
The radial  decrease
in the density of the relativistic electrons
from the position of the shock
at  3 cm  can be obtained
by assuming a diffusive process
and the resulting  intensity of synchrotron emission
can be calculated as the
integral along the line of sight.
Here, we considered the intensity profiles
which arise
from a 3D mathematical diffusion
with constant diffusion coefficient
in the  optically
thin case, see  Fig.  \ref{pn_cut}.
In this  process the  particles can be accelerated
by the Fermi I mechanism at the shock position
in a region of thickness $\ll\, (c-b) \, and \, (b-a) $.
In this case the  theoretical profile as given
in Fig.  \ref{pn_cut} toward the external region
is  concave up.
A second model
with  diffusion in  1D with a step length corresponding to
the relativistic  electron  gyro-radius
is  also analyzed, see the concentration profile
given in Fig.  \ref{soluz_1d} and the
intensity profile given in Fig.  \ref{cut_triangolo}.
These diffusive processes  allow to build up
a  theoretical 2D map  of  the shell-like intensity
, see Fig.  \ref{pnbri}.
A third model analyzes  the radiation from a
shell  with constant density of  emission,
see Figure \ref{ring_cut}.
This is the  simplest model which  produces
an  "U" profile  in the cut
of the intensity
which toward the external region
is  concave down.
In this case the  accelerating  mechanism
can be the Fermi II mechanism characterized by
multiple collisions  in a shell having  thickness  $b-a$.

A model based on the conversion into radiation
of the flux of kinetic energy
explains
the observed decrease in flux at  $6\,cm$, see
equation \ref{fluxtimeastro}
and Figure~\ref{flux} .
This means  that a direct  conversion of the flux
of kinetic energy which varies with time is
a  realistic  model.
Due to the fact that the  observed profile
in intensity is concave up , see Figure \ref{pn_cut} ,
a direct acceleration through the Fermi I mechanism
in thin region around the shock is an acceptable model.

\end{document}